\documentclass[aps, prl, twocolumn, a4paper, reprint, floatfix]{revtex4-1}

\usepackage[utf8]{inputenc}
\usepackage{epsfig}
\usepackage[T1]{fontenc}
\usepackage{t1enc}
\usepackage{graphicx}
\usepackage{amssymb}
\usepackage{amsmath}
\usepackage{relsize}
\usepackage{overpic}
\usepackage{hyperref}

\usepackage[normalem]{ulem}

\usepackage{pdfpages}

\newcommand{\avg}[1]{{\left<#1\right>}}

\usepackage{mathtools}
\def\multiset#1#2{\ensuremath{\left(\kern-.3em\left(\genfrac{}{}{0pt}{}{#1}{#2}\right)\kern-.3em\right)}}

\begin{document}

\title{Parsimonious module inference in large networks}

\author{Tiago P. Peixoto}
\email{tiago@itp.uni-bremen.de}
\affiliation{Institut f\"{u}r Theoretische Physik, Universit\"{a}t Bremen, Hochschulring 18, D-28359 Bremen, Germany}

\pacs{89.75.Hc, 02.50.Tt, 89.70.Cf}

\begin{abstract}
  We investigate the detectability of modules in large networks when the
  number of modules is not known in advance. We employ the minimum
  description length (MDL) principle which seeks to minimize the total
  amount of information required to describe the network, and avoid
  overfitting. According to this criterion, we obtain general bounds on
  the detectability of any prescribed block structure, given the number
  of nodes and edges in the sampled network. We also obtain that the
  maximum number of detectable blocks scales as $\sqrt{N}$, where $N$ is
  the number of nodes in the network, for a fixed average degree
  $\avg{k}$. We also show that the simplicity of the MDL approach yields
  an efficient multilevel Monte Carlo inference algorithm with a
  complexity of $O(\tau N\log N)$, if the number of blocks is unknown,
  and $O(\tau N)$ if it is known, where $\tau$ is the mixing time of the
  Markov chain. We illustrate the application of the method on a large
  network of actors and films with over $10^6$ edges, and a
  dissortative, bipartite block structure.
\end{abstract}

\maketitle

The detection of modules --- or \emph{communities} --- is one of the
most intensely studied problems in the recent literature of network
systems~\cite{fortunato_community_2010, newman_communities_2011}. The
use of generative models for this purpose, such as the stochastic
blockmodel
family~\cite{holland_stochastic_1983,fienberg_statistical_1985,
  faust_blockmodels:_1992, anderson_building_1992,
  hastings_community_2006, garlaschelli_maximum_2008,
  newman_mixture_2007, reichardt_role_2007, hofman_bayesian_2008,
  bickel_nonparametric_2009, guimera_missing_2009,
  karrer_stochastic_2011,ball_efficient_2011, reichardt_interplay_2011,
  decelle_inference_2011, decelle_asymptotic_2011, zhu_oriented_2012,
  baskerville_spatial_2011}, has been gaining increasing
attention. This approach contrasts drastically with the majority of
other methods thus far employed in the field (such as modularity
maximization~\cite{newman_finding_2004}), since not only it is derived
from first-principles, but also it is not restricted to purely
\emph{assortative} and undirected community structures. However, most
inference methods used to obtain the most likely blockmodel assume that
the number of communities is known in
advance~\cite{karrer_stochastic_2011,decelle_asymptotic_2011,
zhao_consistency_2011,moore_active_2011,
chen_fitting_2012,zhang_comparative_2012}. Unfortunately, in most
practical cases this quantity is completely unknown, and one would like
to infer it from the data as well. Here we explore a very efficient way
of obtaining this information from the data, known as the \emph{minimum
description length principle} (MDL)~\cite{grunwald_minimum_2007,
  rissanen_information_2010}, which predicates
that the best choice of model which fits a given data is the one which
most \emph{compresses it}, i.e. minimizes the total amount of
information required to describe it. This approach has been introduced
in the task of blockmodel inference in
Ref.~\cite{rosvall_information-theoretic_2007}. Here we generalize it to
accommodate an arbitrarily large number of communities, and to obtain
general bounds on the detectability of arbitrary community
structures. We also show that, according to this criterion, the maximum
number of detectable blocks scales as $\sqrt{N}$, where $N$ is the
number of nodes in the network. Since the MDL approach results in a
simple penalty on the log-likelihood, we use it to implement an
efficient multilevel Monte Carlo algorithm with an overall complexity of
$O(\tau N\log N)$, where $\tau$ is the average mixing time of the Markov
chain, which can be used to infer arbitrary block structures on very
large networks.

\emph{The model ---} The stochastic blockmodel ensemble is composed of
graphs with $N$ nodes, each belonging to one of $B$ blocks, and the
number of edges between nodes of blocks $r$ and $s$ is given by the
matrix $e_{rs}$ (or twice that number if $r=s$). The
\emph{degree-corrected} variant~\cite{karrer_stochastic_2011} further
imposes that each node $i$ has a degree given by $k_i$, where the set
$\{k_i\}$ is an additional parameter set of the model. The directed
version of both models is analogously defined, with $e_{rs}$ becoming
asymmetric, and $\{k^-_i\}$ together with $\{k^+_i\}$ fixing the in- and
out-degrees of the nodes, respectively. These ensembles are
characterized by their microcanonical entropy $\mathcal{S} = \ln\Omega$,
where $\Omega$ is the total number of network
realizations~\cite{bianconi_entropy_2009}. The entropy can be computed
analytically in both cases~\cite{peixoto_entropy_2012},
\begin{equation}\label{eq:st}
  \mathcal{S}_t \cong E - \frac{1}{2} \sum_{rs}e_{rs}\ln\left(\frac{e_{rs}}{n_rn_s}\right),
\end{equation}
for the traditional blockmodel ensemble and,
\begin{equation}\label{eq:sc}
  \mathcal{S}_c \cong -E -\sum_kN_k\ln k! - \frac{1}{2} \sum_{rs}e_{rs}\ln\left(\frac{e_{rs}}{e_re_s}\right),
\end{equation}
for the degree corrected variant, where in both cases
$E=\sum_{rs}e_{rs}/2$ is the total number of edges, $n_r$ is the number
of nodes which belong to block $r$, and $N_k$ is the total number of
nodes with degree $k$, and $e_r=\sum_se_{rs}$ is the number of
half-edges incident on block $r$. The directed case is
analogous~\cite{peixoto_entropy_2012} (see Supplemental Material for an
overview).

The detection problem consists in obtaining the block partition
$\{b_i\}$ which is the most likely, when given an unlabeled network $G$,
where $b_i$ is the block label of node $i$. This is done by maximizing
the log-likelihood $\ln\mathcal{P}$ that the network $G$ is observed,
given the model compatible with a chosen block partition. Since we have
simply $\mathcal{P} = 1/\Omega$, maximizing $\ln\mathcal{P}$ is
equivalent to minimize the entropy $\mathcal{S}_{t/c}$, which is the
language we will use henceforth. Entropy minimization is well-defined,
but only as long as the total number of blocks $B$ is known
beforehand. Otherwise, the optimal value of $\mathcal{S}_{t/c}$ becomes
a strictly decreasing function of $B$. Thus, simply minimizing the
entropy will lead to the trivial $B=N$ partition, and the block matrix
$e_{rs}$ becomes simply the adjacency matrix. A principled way of
avoiding such overfitting is to consider the total amount of information
necessary to describe the data, which includes not only the entropy of
the fitted model, but also the information necessary to describe the
\emph{model itself}. This quantity is called the \emph{description
length}, and for the stochastic blockmodel ensemble it is given by
\begin{equation}\label{eq:sigma}
  \Sigma_{t/c} = \mathcal{S}_{t/c} + \mathcal{L}_{t/c},
\end{equation}
where $\mathcal{L}_{t/c}$ is the information necessary to describe the
model via the $e_{rs}$ matrix and the block assignments $\{b_i\}$. The
minimum value of $\Sigma_{t/c}$ is an upper bound on the total amount of
information necessary to describe a given network to an observer lacking
any \emph{a priori}
information~\cite{rosvall_information-theoretic_2007}. Therefore, the
best model chosen is the one which \emph{best compresses} the data,
which amounts to an implementation of Occam's Razor. For the specific
problem at hand, it is easy to compute $\mathcal{L}_{t/c}$. The $e_{rs}$
matrix can be viewed as the adjacency matrix of a multigraph with $B$
nodes and $E$ edges, where the blocks are the nodes and self-loops are
allowed. The total number of $e_{rs}$ matrices is then simply
$\multiset{\multiset{B}{2}}{E}$~\footnote{Where $\multiset{n}{k} =
{n+k-1\choose k}$ is the number of $k$-combinations with repetitions
from a set of size $n$.}. The total number of block partitions is $B^N$.
Assuming no prior information on the model, we obtain $\mathcal{L}_t$ by
multiplying these numbers and taking the logarithm,
\begin{equation}\label{eq:lt}
  \mathcal{L}_t \cong Eh\left(\frac{B(B+1)}{2E}\right) + N\ln B
\end{equation}
where $h(x) = (1+x)\ln (1+x) -x\ln x $, and $E \gg 1$ was assumed.  Note
that Eq.~\ref{eq:lt} is not the same as the expression derived in
Ref.~\cite{rosvall_information-theoretic_2007}, which is obtained by
taking the limit $E \gg B^2$, in which case we have $\mathcal{L}_t
\approx \frac{B(B+1)}{2}\ln E + N\ln B$~\footnote{The value of
$\mathcal{L}_t$ was computed in
Ref.~\cite{rosvall_information-theoretic_2007} as the number of
symmetric matrices with entry values from $0$ to $E$, without the
restriction that the sum must be exactly $2E$, which is accounted for in
Eq.~\ref{eq:lt}. If $E\gg B^2$ this restriction can be neglected, but
not otherwise.}. We do not take this limit~\emph{a priori}, since, as we
show below, block sizes up to $B_{\text{max}} \sim \sqrt{E}$ can in
principle be detected from empirical data.  For the degree-corrected
variant, we still need to describe the degree sequence of the network,
hence
\begin{equation}\label{eq:lc}
\mathcal{L}_c = \mathcal{L}_t - N\sum_kp_k\ln p_k,
\end{equation}
where $p_k$ is the fraction of nodes with degree $k$.  Note that for the
directed case we need simply to replace $B(B+1)/2 \to B^2$ and $k\to
(k^-,k^+)$ in the equations above.

\begin{figure}
  \begin{minipage}{0.49\columnwidth}
    (a)\begin{minipage}[b]{\columnwidth}
      \hspace{-2.6em}
        \begin{minipage}[b]{0.5\columnwidth}
          \centering
          \includegraphics[width=\columnwidth]{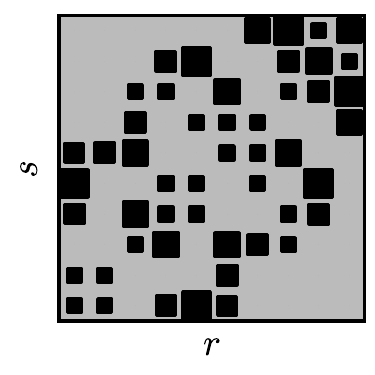}\\
          \smaller[2] {}\hspace{0.6em} Planted
          \vspace{0.6em}
        \end{minipage}
        \hspace{-0.8em}
        \begin{minipage}[b]{0.58\columnwidth}
          \centering
          \begin{minipage}[b]{0.45\columnwidth}
            \centering
            \includegraphics[width=\columnwidth]{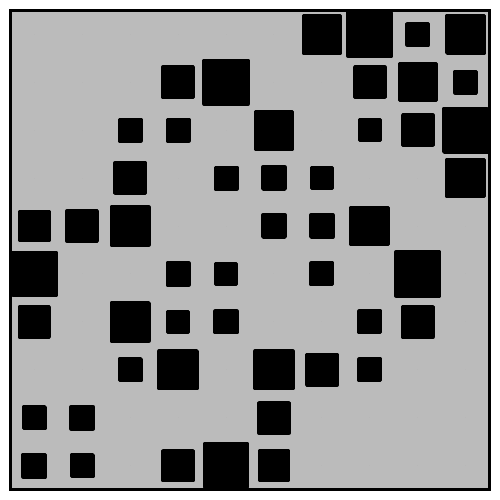}
            \smaller[4] $\avg{k} = 15$
          \end{minipage}
          \begin{minipage}[b]{0.45\columnwidth}
            \centering
            \includegraphics[width=\columnwidth]{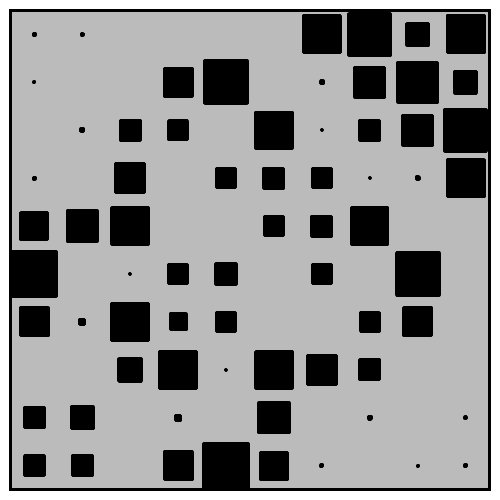}
            \smaller[4] $\avg{k} = 6$
          \end{minipage}
          \begin{minipage}[b]{0.45\columnwidth}
            \centering
            \includegraphics[width=\columnwidth]{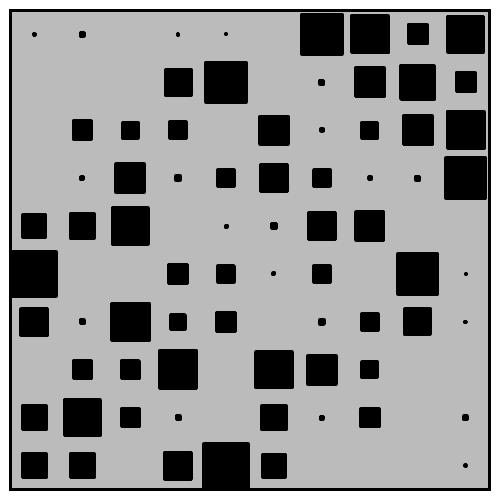}
            \smaller[4] $\avg{k} = 5$
          \end{minipage}
          \begin{minipage}[b]{0.45\columnwidth}
            \centering
            \includegraphics[width=\columnwidth]{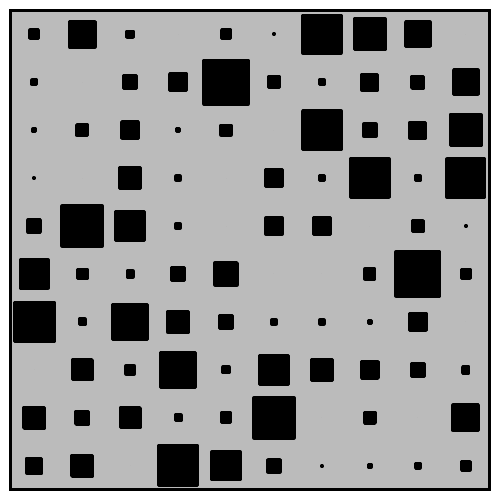}
            \smaller[4] $\avg{k} = 4$
          \end{minipage}
          \smaller[2] Inferred
        \end{minipage}
        \vspace{0.3em}
      \end{minipage}
  \end{minipage}
  \begin{minipage}{0.49\columnwidth}
    \begin{overpic}[unit=1mm, width=\columnwidth, trim=0.1cm 0 0 0, clip]{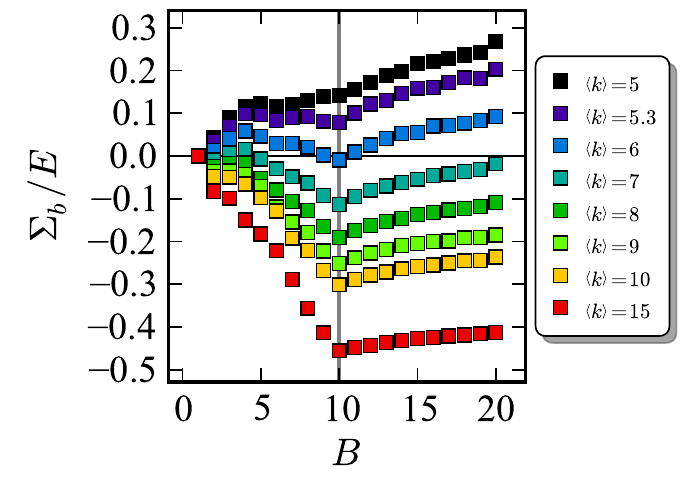}
      \put(3,8){(b)}
    \end{overpic}
  \end{minipage}
  \begin{minipage}{0.494\columnwidth}
    \begin{overpic}[unit=1mm, width=\columnwidth, trim=0.248cm 0 0.3cm 0, clip]{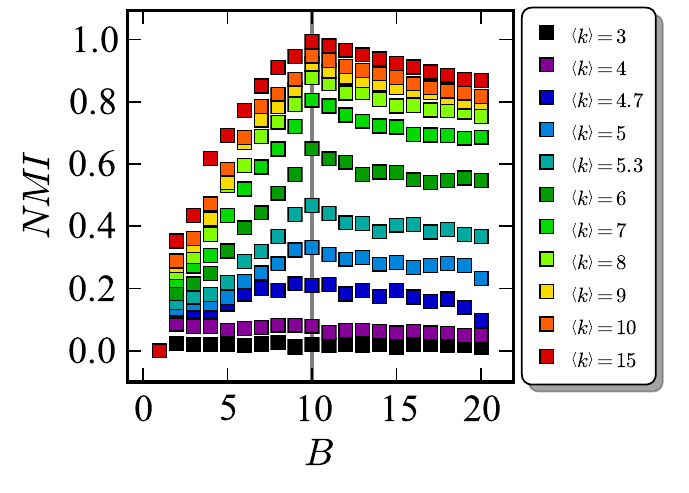}
      \put(3,8){(c)}
    \end{overpic}
  \end{minipage}
  \begin{minipage}{0.494\columnwidth}
    \begin{overpic}[unit=1mm, width=\columnwidth, trim=0.24cm 0 0.05cm 0, clip]{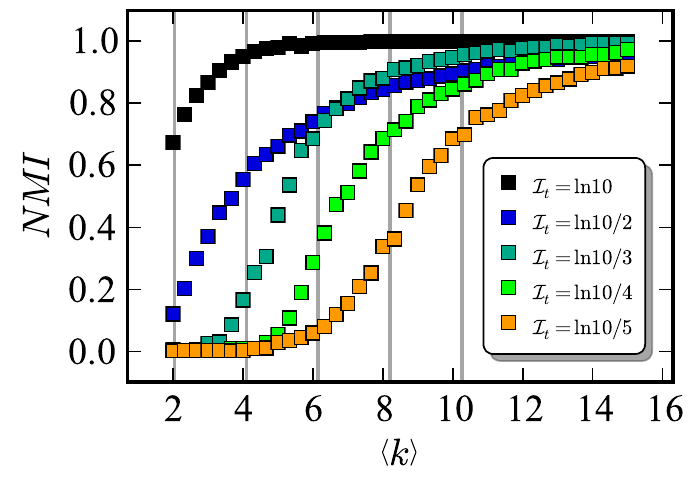}
      \put(3,8){(d)}
    \end{overpic}
  \end{minipage}
  \caption{
    \label{fig:example}(a) Prescribed block structure with $B=10$ and
    $\mathcal{I}_{t} = \ln B / 6$, together with inferred parameters for
    different $\avg{k}$; (b) Description length $\Sigma_b/E$ for
    different $B$ and $\avg{k}$, for networks sampled from (a). The
    vertical line marks the position of the global minimum; (c) NMI
    between the true and inferred partitions, for the same networks as
    in (b); (d) Same as (b) for different $\avg{k}$ and prescribed block
    structures. The grey lines correspond to the threshold of
    Eq.~\ref{eq:kc}. In all cases we have $N=10^4$.}
\end{figure}

\emph{MDL bound on detectability ---} The difference $\Sigma_b \equiv
\Sigma_{t/c} - \Sigma_{t/c}|_{B=1}$ of the description length of a graph
with some block structure and a random graph with $B=1$ can be written
as
\begin{equation}\label{eq:dl_b}
  \Sigma_b = Eh\left(\frac{B(B+1)}{2E}\right) + N\ln B - E\mathcal{I}_{t/c},
\end{equation}
with $\mathcal{I}_t = \sum_{rs}m_{rs}\ln(m_{rs}/w_rw_s)$ and
$\mathcal{I}_c = \sum_{rs}m_{rs}\ln(m_{rs}/m_rm_s)$, where $m_{rs} =
e_{rs}/2E$ and $w_r=n_r/N$ (and equivalently for directed graphs, with
$B(B+1)/2 \to B^2$). We note that $\mathcal{I}_{t/c}\in[0, \ln B]$.  If
for any given graph we have $\Sigma_b > 0$, the inferred block structure
will be discarded in favor of the simpler fully random $B=1$
model. Therefore the condition $\Sigma_b < 0$ yields a limit on the
detectability of prescribed block structures according to the MDL
criterion. For the special case where $E \gg B^2$, this inequality
translates to a more convenient form,
\begin{equation}\label{eq:kc}
  \avg{k} > \frac{2\ln B}{\mathcal{I}_{t/c}}.
\end{equation}
The directed case is analogous, with $2\ln B \to \ln B$ replaced in the
equation above.

\begin{figure}
  \begin{overpic}[unit=1mm, width=0.49\columnwidth, trim=0.25cm 0 0.09cm 0, clip]{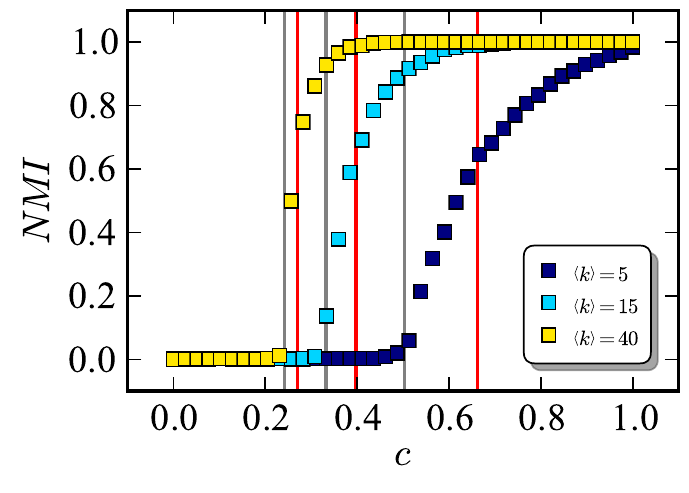}
    \put(3,8){(a)}
  \end{overpic}
  \begin{overpic}[unit=1mm, width=0.49\columnwidth, trim=0.12cm 0 0.05cm 0, clip]{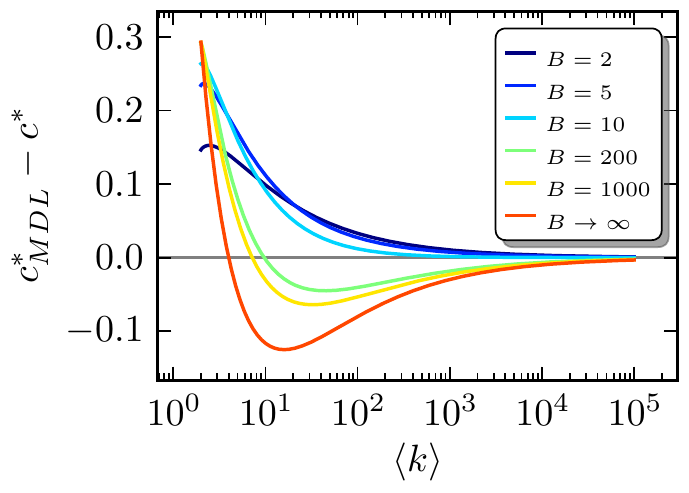}
    \put(3,8){(b)}
  \end{overpic} \caption{\label{fig:planted}(a) NMI between the true and
  inferred partitions for PP samples with $B=10$ as a function of $c$
  for different $\avg{k}$. The grey (red) lines correspond to the
  threshold $c^*$ of Ref.~\cite{decelle_inference_2011}
  ($c^*_{\text{MDL}}$ given by Eq.~\ref{eq:kc}); (b) Difference between
  $c^*_{\text{MDL}}$ and $c^*$, for different $\avg{k}$ and $B$.}
\end{figure}

\emph{Partial detectability and parsimony ---} The condition $\Sigma_b <
0$ is not a statement on the absolute detectability of a given model,
only to what extent the extracted information (if any) can be used to
\emph{compress the data}. Although these are intimately related, the MDL
criterion is based on the idea of perfect (or \emph{lossless})
compression, and thus corresponds simply to a condition necessary (but
not sufficient) for the \emph{perfect} recoverability of the model
parameters from the data. Perfect inference, however, is only possible
in the asymptotically dense case
$\avg{k}\to\infty$~\cite{decelle_asymptotic_2011}, and in practice one
always has some amount of uncertainty. Therefore it remains to be
determined how practical is the parsimony limit derived from MDL to
establish a noise threshold on empirical data. In Fig.~\ref{fig:example}
is shown an example of a block structure with $B=10$ and $\mathcal{I}_t
= \ln B / 6$. In Fig.~\ref{fig:example}b is shown the minimum of
$\Sigma_b/E$ as function of $B$, for sampled networks with different
$\avg{k}$, obtained with the Monte Carlo algorithm described below. If
$\avg{k}$ is large enough ($\avg{k}
> 6$, according to Eq.~\ref{eq:kc}), the minimum of $\Sigma_b$ is
clearly at the correct $B=10$ value, and as is show in
Fig.~\ref{fig:example}b this is exactly where the normalized mutual
information (NMI)~\footnote{NMI here is defined as $2I(X,Y)/(H(X) +
H(Y))$, where $I(X,Y)$ is the mutual information between $X$ and $Y$,
and $H(X)$ is the entropy of $X$.}  between the known and inferred
partition is the largest. However, for $\avg{k} < 6$ the minimum of
$\Sigma_b$ is no longer at $B=10$, and instead it is at
$B=1$. Nevertheless, the overlap with the correct partition is overall
positive and is still is the largest at $B=10$, so the correct partition
is to some extent detectable, but the MDL criterion rejects it. By
experimenting with different planted block structures (see
Fig.~\ref{fig:example}d), one observes that the MDL threshold lies very
close to the parameter region where inferred partition is no longer well
correlated with the true partition.  This comparison can be made in more
detail by considering the special case known as the \emph{planted
partition model} (PP)~\cite{condon_algorithms_2001}, which imposes a
diagonal block structure given by $m_{rr} = c / B$, $m_{rs} = (1-c) /
B(B-1)$ for $r\ne s$, and $w_r=1/B$, and $c\in[0, 1]$ is a free
parameter. In this case it can be shown that even partial inference is
only possible if $\avg{k} >
((B-1)/(cB-1))^2$~\cite{decelle_inference_2011, decelle_asymptotic_2011,
mossel_stochastic_2012, nadakuditi_graph_2012}, otherwise no information
at all on the original model can be extracted~\footnote{In
\cite{decelle_inference_2011, nadakuditi_graph_2012} this threshold was
expressed with a different notation, as a function of
$c_{\text{in}}=\avg{k}cB$ and $c_{\text{out}}=\avg{k}(1-c)B/(B-1)$,
instead of $c$. It is also common to express such transitions as a
function of $k_{in} = c \avg{k}$ and $k_{out} = (1-c) \avg{k}$, or the
mixing parameter $\mu = k_i^{\text{out}} / (k_i^{\text{in}} +
k_i^{\text{out}})$~\cite{lancichinetti_benchmark_2008,lancichinetti_community_2009}.
For the PP model, we have simply $\mu\simeq 1 - c$, for sufficiently
large degrees.}. For smaller values of $B$, this bound is higher than
Eq.~\ref{eq:kc} for this model (where we have $\mathcal{I}_{t/c} =
c\ln(Bc)+(1-c)\ln(B(1-c)/(B-1)$), which means that there is a region of
parameters where the MDL criterion discards potentially useful (albeit
clearly noisy) information (see Fig.~\ref{fig:planted}a). Interestingly,
however, for larger values of $B$, the MDL criterion will most often
result in \emph{lower} bounds (see Fig.~\ref{fig:planted}b), meaning
that whatever partial information which can be recovered from the model
will \emph{not} be discarded. For $B\to\infty$ we have
$c^*_{\text{MDL}}\simeq 2/\avg{k}$ and $c^*\simeq 1 / \sqrt{\avg{k}}$,
and thus $c^*_{\text{MDL}} < c^*$ for $\avg{k} > 4$~\footnote{Here we
impose $E\gg B^2$ first, and $B\to\infty$ later.}. Therefore, so far as
the PP model serves as a good representation of more general block
structures, one should not expect excessive parsimony from MDL, at least
for sufficiently large values of $B$.

\emph{The largest detectable value of B ---} The MDL approach imposes an
intrinsic constraint on the maximum value of $B$ which can be detected,
$B_{\text{max}}$, given a network size and density. This can be obtained
by minimizing $\Sigma_b$ over all possible block structures with a given
$B$, which is obtained simply by replacing $\mathcal{I}_{t/c}$ by its
maximum value $\ln B$ in Eq.~\ref{eq:dl_b},
\begin{equation}\label{eq:dl_b_min}
  \Sigma'_b = Eh\left(\frac{B(B+1)}{2E}\right) - (E-N)\ln B.
\end{equation}
Eq.~\ref{eq:dl_b_min} is a strictly convex function on $B$. This means
there is a global minimum $\Sigma'_b|_{B=B_{\text{max}}}$ given uniquely
by $N$ and $E$.  It is easy to see that even if the prescribed block
structure with some $B>B_{\text{max}}$ has minimal entropy
(i.e. $\mathcal{I}_{t/c} = \ln B$), alternative partitions with $B'<B$
blocks (obtained by merging blocks such that $\mathcal{I}'_{t/c} = \ln
B'$) will necessarily possess a smaller $\Sigma'_b$. Imposing $\partial
\Sigma'_b / \partial B = 0$, one obtains $B_{\text{max}} \cong
\mu(\avg{k})\sqrt{E}$, with $\mu(\avg{k})$ being the solution of
$\mu\ln(2/\mu^2 + 1) - (1-1/\avg{k})/\mu = 0$ [for the directed case we
make $2/\mu^2\to 1/\mu^2$ and $1/\avg{k} \to 2/\avg{k}$]. Therefore,
according to the MDL criterion, the maximum number of blocks which is
detectable scales as $B_{\text{max}}\sim\sqrt{N}$ for a fixed value of
$\avg{k}$. This is consistent with detectability analysis in
Ref.~\cite{choi_stochastic_2012} for traditional blockmodel variant,
which showed by other means that the model parameters can only be
recovered if $B$ does not scale faster than $\sqrt{N}$. Note that this
means that the limit $E\gg B^2$ cannot be taken \emph{a priori} when
inferring from empirical data, and hence the value of $\mathcal{L}_t$
computed in Ref.~\cite{rosvall_information-theoretic_2007} needs to be
replaced with Eq.~\ref{eq:lt} in the general case.

The limit $B_{\text{max}} \propto \sqrt{E}$ is very similar to the
so-called ``resolution limit'' of community detection via modularity
optimization~\cite{fortunato_resolution_2007}, which is
$B^Q_{\text{max}} = \sqrt{E}$. These two limits, however, have different
interpretations: The value of $B^Q_{\text{max}}$ arises simply from the
definition of modularity, which can be to some extent alleviated (but
not entirely avoided) by properly modifying the modularity function with
scale
parameters~\cite{reichardt_statistical_2006,kumpula_limited_2007,arenas_analysis_2008,
lancichinetti_limits_2011,
xiang_limitation_2011,ronhovde_local_2012}. On the other hand the value
of $B_{\text{max}}$ has a more fundamental character, and corresponds to
the situation where knowledge of the complete block structure is no
longer the best option to compress the data. This value can be improved
only if any \emph{a priori} information is known which leads to a
smaller class of models to be inferred, and hence smaller
$\mathcal{L}_t$. In general, if we have $\mathcal{L}_t = Ef(B^\alpha/E)
+ N\ln B$, where $f(x)$ is any (differentiable) function, performing the
same analysis as above leads to $B_{\text{max}}=(\mu(k)E)^{1/\alpha}$, with $\alpha
f'(\mu)\mu + 2/\avg{k}-1=0$. However, it should also be noted that if the existing
block structure is locally dense (i.e. $e_{rs} \sim n_rn_s$), as the
union of $B$ complete graphs considered
in~\cite{fortunato_resolution_2007}, the expressions in Eqs.~\ref{eq:st}
and~\ref{eq:sc} are no longer valid, and will overestimate the
entropy. Using the correct entropy (Eqs.~5 and 9
in~\cite{peixoto_entropy_2012}) will lead to an improved
resolution. Unfortunately, for the dense case, the entropy for the
degree-corrected variant cannot be computed in a closed
form~\cite{peixoto_entropy_2012}.

\begin{figure}[t]
  \begin{minipage}{0.49\columnwidth}
    \begin{center}
      \includegraphics[width=\columnwidth,trim=0.27cm 0cm 0.1cm 0cm, clip]{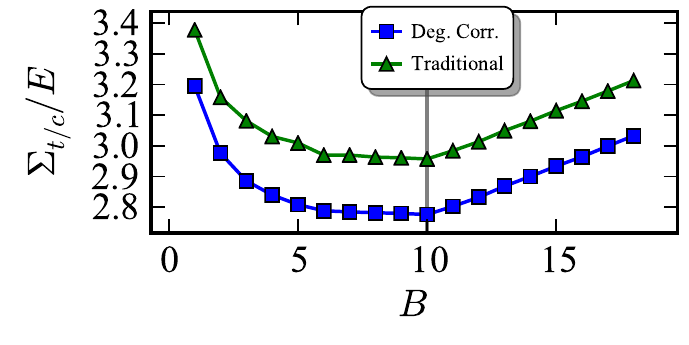}
      \begin{overpic}[unit=1mm, width=0.8\columnwidth, trim=0 0 0 1cm, clip]{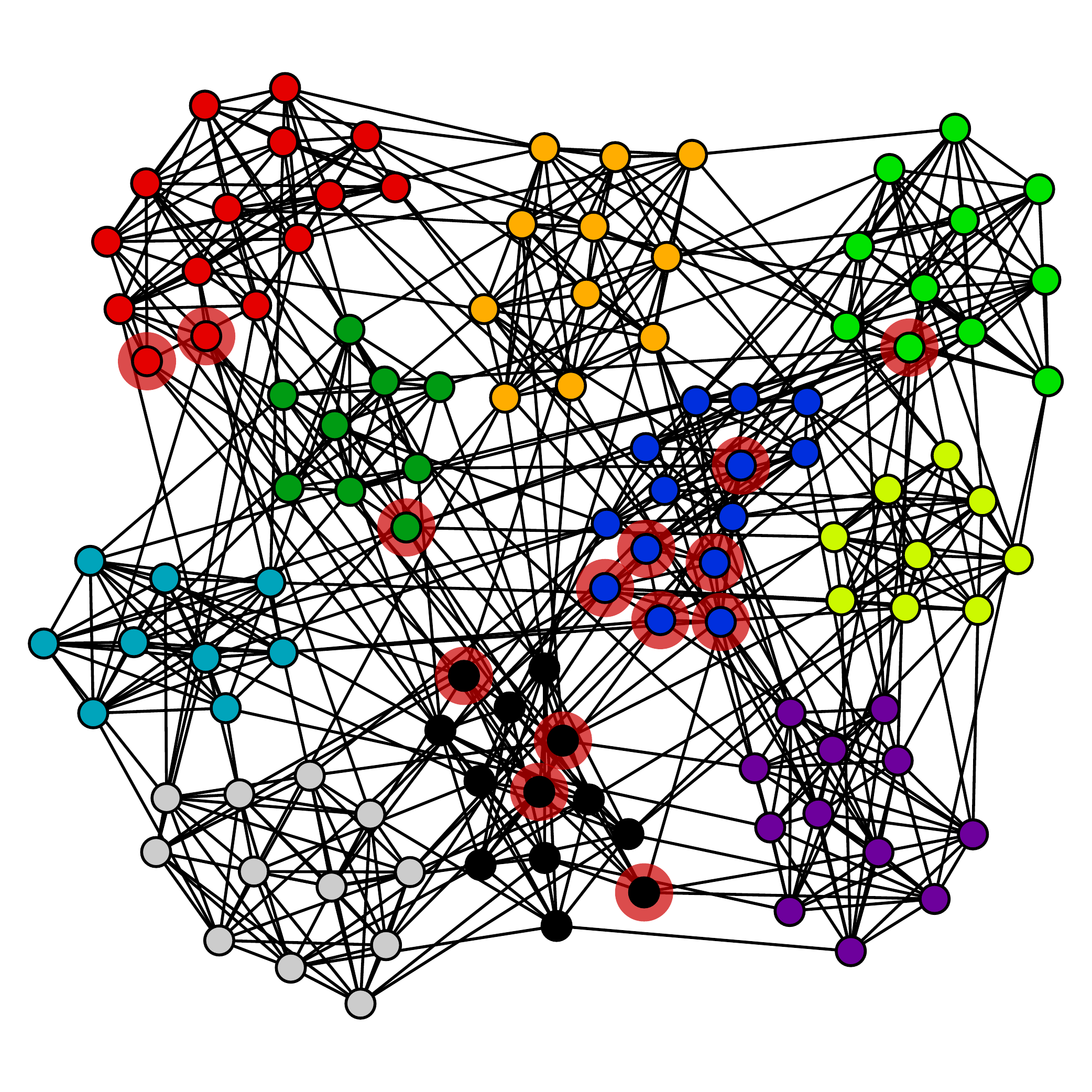}
        \put(-5,8){(a)}
      \end{overpic}
    \end{center}
  \end{minipage}
  \begin{minipage}{0.49\columnwidth}
    \begin{center}
      \includegraphics[width=\columnwidth,trim=0.27cm 0cm 0.1cm 0cm, clip]{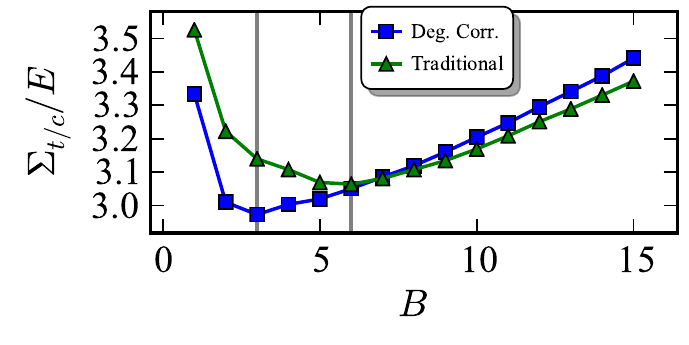}
      \begin{overpic}[unit=1mm, width=0.8\columnwidth, trim=0 0 0 1cm, clip]{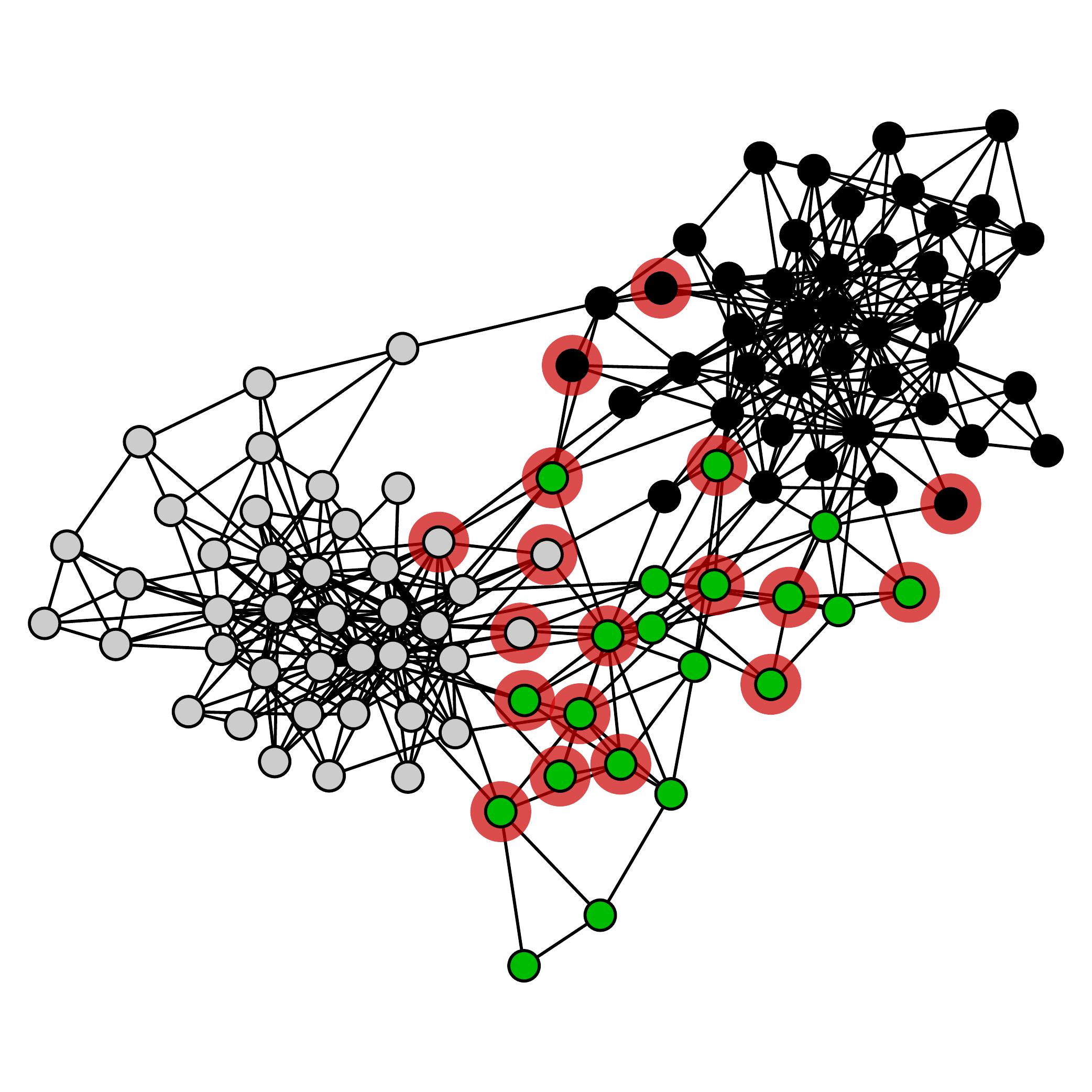}
        \put(3,8){(b)}
      \end{overpic}
    \end{center}
  \end{minipage}
 \caption{\label{fig:emprical-small}\emph{Top:} Value of
  $\Sigma_b/E$ for both blockmodel variants as a function of $B$ for (a)
  the American football network of~\cite{girvan_community_2002} (with
  the corrections described
  in~\cite{evans_clique_2010,evans_american_2012}) and (b) the political
  books network of~\cite{krebs_political_????}. \emph{Bottom:} Inferred
  partitions with the smallest $\Sigma_b$. Nodes circled in red do not
  match the known partitions.}
\end{figure}
\begin{figure}[ht]
  \begin{minipage}{1\columnwidth}
    \begin{minipage}[c]{0.35\columnwidth}
      \includegraphics[width=1\columnwidth]{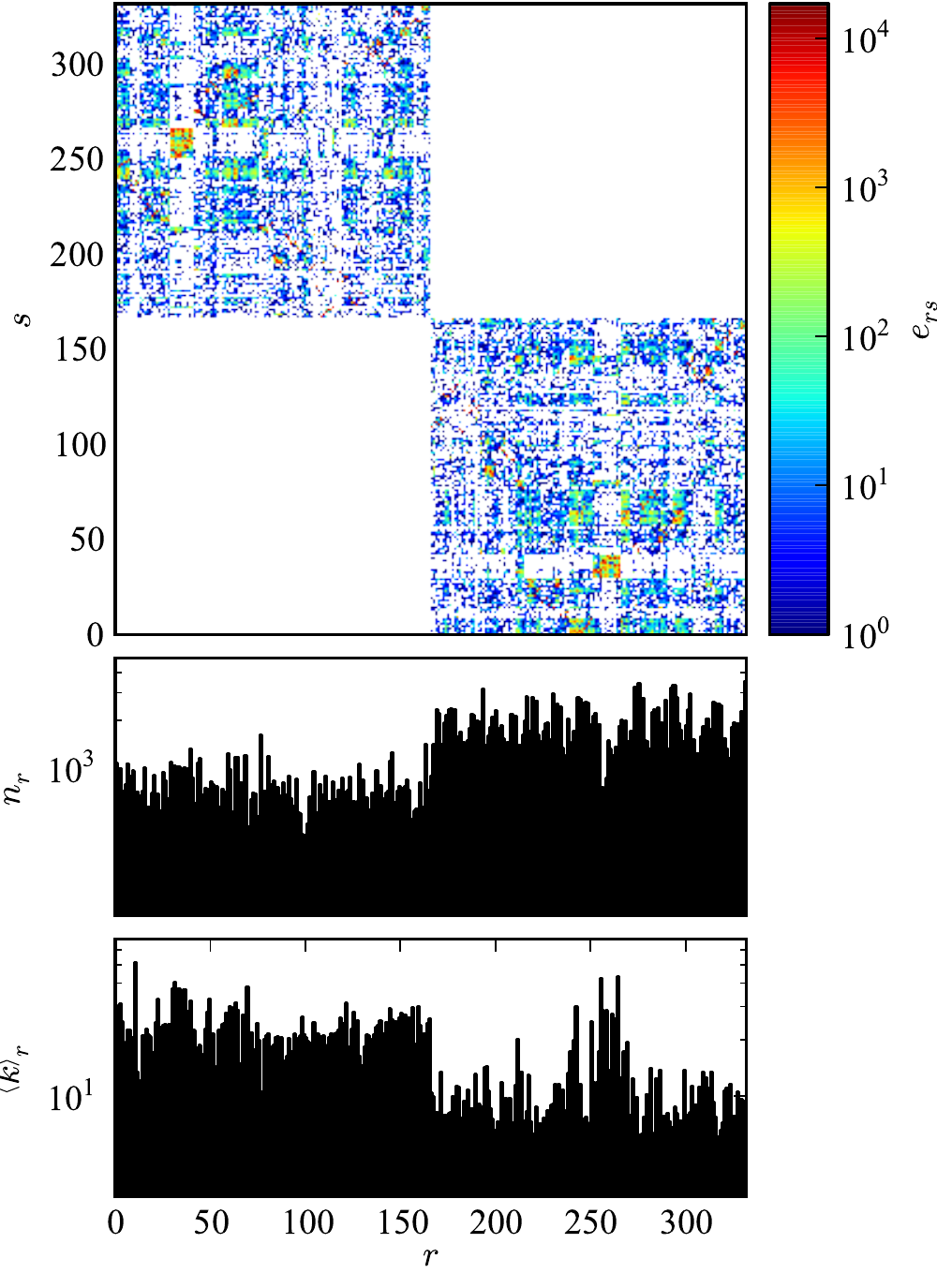}
    \end{minipage}
    \hspace{1.5em}
    \begin{minipage}[c]{0.49\columnwidth}
      \includegraphics[width=1\columnwidth, trim=0.7cm 0 0.8cm 0, clip]{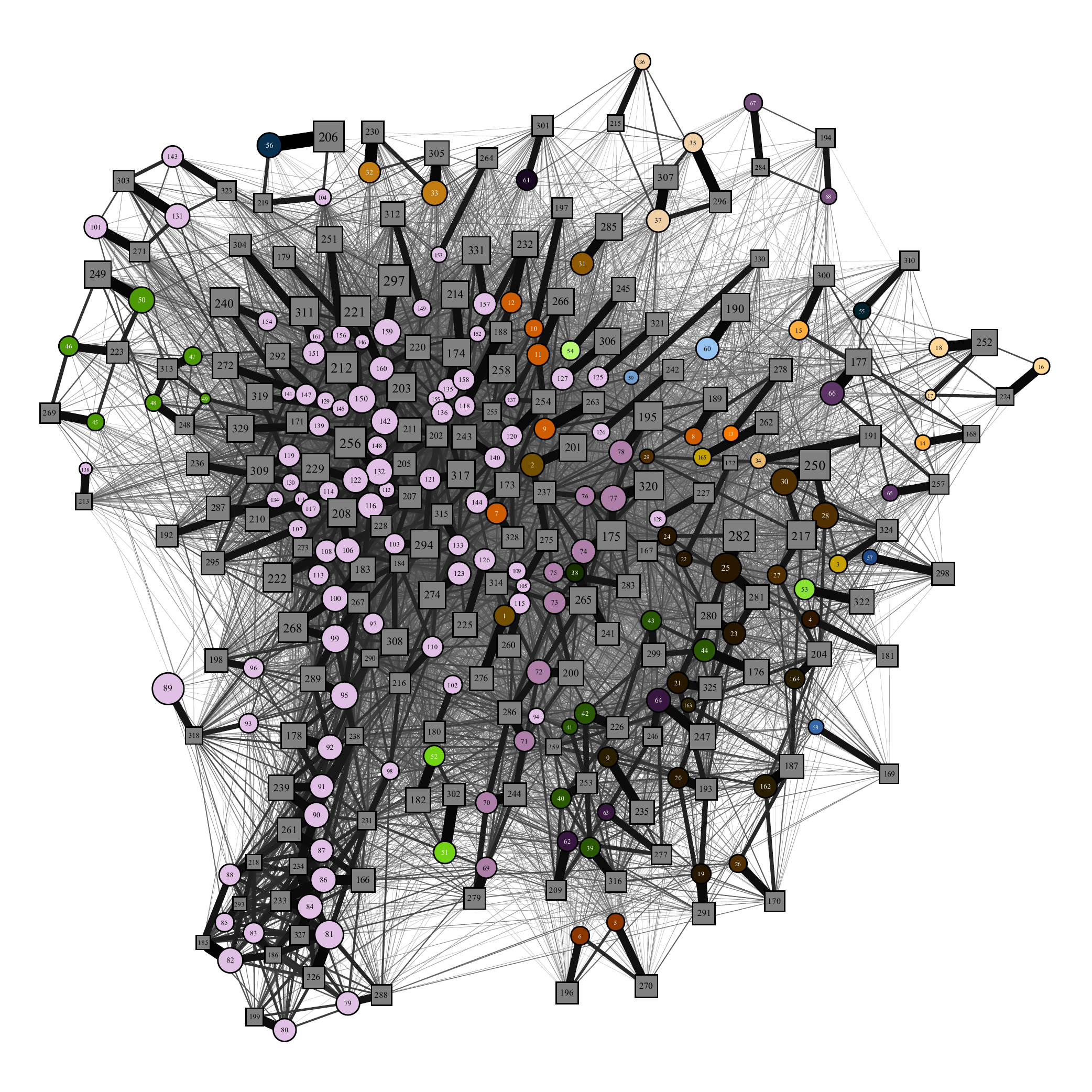}
    \end{minipage}
  \end{minipage} \caption{\label{fig:imdb}\emph{Left:} Inferred block
  structure for the IMDB network, with $N=372,787$, $E=1,812,657$ and
  $B=332$, according to the MDL criterion, and the degree-corrected
  stochastic blockmodel. \emph{Right:} Circles correspond to film
  blocks, and squares to actors. The node colors correspond to the
  countries of production. See Supplemental Material for more details.}
\end{figure}

\emph{Detection algorithm ---} For a fixed $B$, the best partition can
be found by minimizing $\mathcal{S}_{t/c}$, via well-established methods
such as Markov chain Monte Carlo (MCMC), using the the
Metropolis-Hastings
algorithm~\cite{metropolis_equation_1953,hastings_monte_1970}. However,
a naïve implementation based on fully random block membership moves can
be very slow. We found that the performance can be drastically improved
by using local information and current knowledge of the partially
inferred block structure, simply by proposing moves $r\to s$ with a
probability $p(r\to s|t) \propto e_{ts}
+ 1$, where $t$ is the block label of a randomly chosen neighbor of the
node being moved. Each sweep of this algorithm can be performed in
$O(E)$ time, independent of $B$ (see Supplemental Material). Having
obtained the minimum of $\mathcal{S}_{t/c}$, the best value of $B$ is
obtained via an independent one-dimensional minimization of $\Sigma_b$,
using a Fibonacci search~\cite{press_numerical_2007}, based on
subsequent bisections of an initial interval which brackets the
minimum. This method finds a local minimum in $O(\ln B_{\text{max}})$
time. The overall number of steps necessary for the entire algorithm is
$O(\tau E \ln B_{\text{max}})$, where $\tau$ is the average mixing time
of the Markov chain. If we have no prior information on
$B_{\text{max}}$, we need to assume $B_{\text{max}}\sim\sqrt{E}$, in
which case the complexity becomes $O(\tau E\ln E)$, or $O(\tau N\ln N)$
for sparse graphs. This compares favorably to minimization strategies
which require the computation of the full marginal probability $\pi^i_r$
that node $i$ belongs to block $r$, such as Belief-Propagation
(BP)~\cite{decelle_inference_2011,
decelle_asymptotic_2011,yan_model_2012}, which results in a larger
complexity of $O(NB^2)$ per sweep (or $O(NB^2l)$ for the
degree-corrected variant, with $l$ being the number of distinct
degrees~\cite{yan_model_2012}), or $O(N^2)$ for $B \sim B_{\text{max}}$.

\emph{Empirical networks ---} The MDL approach yields convincing results
for many empirical networks, as can be seen in
Fig.~\ref{fig:emprical-small}, which shows results for the College
Football network of~\cite{girvan_community_2002} and the Political Books
network of~\cite{krebs_political_????}. In both cases the correct number
blocks is inferred, and the best partition matches reasonably well the
known true values, at least for the degree-corrected variant. Employing
the Monte Carlo algorithm above, results may be obtained for much larger
networks. We show in Fig.~\ref{fig:imdb} the obtained block partition
with the degree-corrected variant for the IMDB network of actors and
films~\footnote{Retrieved from \texttt{http://www.imdb.com/interfaces}},
where a film node is connected to all its cast members. The
bipartiteness of the network is fully reflected in the inferred block
partition, where films and actors always belong to different blocks,
although this has not been imposed \emph{a priori} (something which
would be impossible to obtain with, e.g. modularity
optimization). Besides this role separation, the film blocks are divided
sharply along spatial, temporal and genre lines, and the actor blocks
are closely correlated with such film classes (see Supplemental Material
for a more detailed analysis).

In summary, we showed how minimizing the full description length of
empirical network data enables simple, efficient, unbiased and fully
non-parametric analysis of the large-scale properties of large networks,
for which no \emph{a priori} information is available, while at the same
time providing general bounds on the decodability of arbitrary block
structures from empirical data.

\begin{acknowledgments}
I would like to thank Tim S. Evans for pointing out some corrections to
the American football data, and Laerte B. P. de Andrade for useful
conversations about the IMDB data.
\end{acknowledgments}

\bibliographystyle{apsrev4-1} \bibliography{bib}

\newpage\mbox{}\newpage
\onecolumngrid


\section*{Supplementary material}

# Supplemental Material: Parsimonious Module Inference in Large Networks

Tiago P. Peixoto*

*Institut für Theoretische Physik, Universität Bremen, Hochschulring 18, D-28359 Bremen, Germany*

## I. BLOCKMODEL ENTROPY

We give here a brief overview of the entropies of the various blockmodel variants referenced in the main text. For more details refer to Ref [1]. The entropy of the traditional blockmodel ensemble for the undirected case is

$$\mathcal{S}_t = \frac{1}{2} \sum_{rs} e_{rs} H\left(\frac{e_{rs}}{n_r n_s}\right), \qquad (1)$$

while for the directed case it reads,

$$\mathcal{S}_t^d = \sum_{rs} e_{rs} H\left(\frac{e_{rs}}{n_r n_s}\right), \qquad (2)$$

where $H(x) = -x \ln x - (1-x) \ln(1-x)$ is the binary entropy function. In both cases, $e_{rs}$ is the number of edges from block $r$ to $s$ (or the number of half-edges for the undirected case when $r = s$), and $n_r$ is the number of nodes in block $r$. In the sparse limit, $e_{rs} \ll n_r n_s$, these expressions may be written approximately as,

$$\mathcal{S}_t \cong E - \frac{1}{2} \sum_{rs} e_{rs} \ln\left(\frac{e_{rs}}{n_r n_s}\right), \qquad (3)$$

$$\mathcal{S}_t^d \cong E - \sum_{rs} e_{rs} \ln\left(\frac{e_{rs}}{n_r n_s}\right). \qquad (4)$$

However, since Eqs. 1 and 2 are more general, they should in principle be preferred. On the other hand, Eqs. 3 and 4 have the advantage that they allow the entropy difference $\Delta \mathcal{S}_t$ obtained by changing the block membership of a single node to be computed more easily. For the undirected case, for instance, we may write $\mathcal{S}_t \cong E - \frac{1}{2} \sum_{rs} e_{rs} \ln e_{rs} + \sum_r e_r \ln n_r$, and notice that we need to modify at most $4k$ terms in the first sum and 2 terms in the second if we change the membership of a node with degree $k$ (this also true for the degree-corrected variant below). On the other hand, using Eq. 1 we need to modify a number of terms which is proportional to the number of blocks $B$, which will become costly if it is much larger than typical values of $k$. Thus the extra precision comes at a performance cost, at least when using a Monte Carlo algorithm depending on block membership moves. Therefore, if one is assumes that the sparsity condition $e_{rs} \ll n_r n_s$ is likely to be fulfilled, using Eqs. 3 and 4 can be advantageous.

For the degree-corrected variant with "hard" degree constraints [2] the equivalent expressions are

$$\mathcal{S}_c \cong -E - \sum_k N_k \ln k! - \frac{1}{2} \sum_{rs} e_{rs} \ln\left(\frac{e_{rs}}{e_r e_s}\right), \qquad (5)$$

$$\mathcal{S}_c^d \cong -E - \sum_{k^+} N_{k^+} \ln k^+! - \sum_{k^-} N_{k^-} \ln k^-! - \sum_{rs} e_{rs} \ln\left(\frac{e_{rs}}{e_r^+ e_s^-}\right), \qquad (6)$$

where $e_r = \sum_s e_{rs}$ is the number of half-edges incident on block $r$, and $e_r^+ = \sum_s e_{rs}$ and $e_r^- = \sum_s e_{sr}$ are the number of out- and in-edges adjacent to block $r$, respectively. These expressions are also only valid in the sparse limit, which in this case involves more details of the degree distribution,

$$e_{rs} \frac{\langle k^2 \rangle_r - \langle k \rangle_r}{\langle k \rangle_r^2} \frac{\langle k^2 \rangle_s - \langle k \rangle_s}{\langle k \rangle_s^2} \ll n_r n_s, \qquad (7)$$

---

* tiago@itp.uni-bremen.de

where $\langle k^l \rangle_r = \sum_{i \in r} k_i^l / n_r$ (for the directed case we simply replace $\langle k^l \rangle_r \to \langle (k^+)^l \rangle_r$ and $\langle k^l \rangle_s \to \langle (k^-)^l \rangle_s$ in the equation above). Unfortunately there is no closed-form expression for the entropy outside the sparse limit, unlike the traditional variant. One can derive higher-order corrections which become relevant when the dense limit is approached [1], which will be different for simple and multigraphs, as well as "hard" or "soft" degree constraints, but will break down if the deviation from the sparse case is too strong. This could be specially problematic for networks with a very broad degree distribution without a structural cutoff, for which the values of $\langle k^2 \rangle_r$ are large. In such cases Eqs. 5 and 6 can still be used, but they must be understood as approximations, which may lead to the detection of spurious blocks, which simply reflect the intrinsic dissortative degree-degree correlations [1]. The minimum description length approach presented in the main text may alleviate this problem, since these spurious blocks may end up being rejected if the dissortativity is not too strong. But more care should be taken in such cases, since a more satisfying general methodology is still lacking.

## II. MONTE CARLO INFERENCE

As described in the main text, the Markov chain Monte Carlo (MCMC) inference algorithm consists in using the Metropolis-Hastings algorithm [3, 4], where for each node $i$ one attempts a move $r \to s$, where $r = b_i$ is its current block membership, and accept or reject the move depending on the entropy difference. The main caveat here is how the proposed values of $s$ are chosen. The simplest approach is to chose randomly between all $B$ options, which would lead to a correct algorithm, but with a very slow convergence to the steady state distribution for large values of $B$, since most moves would simply be rejected. Instead, we opt to propose moves which take into account the partial block structure inferred at the current stage of the algorithm and the local neighbourhood of the node being moved: We inspect a random neighbour $j$ of the node $i$ being moved, and obtain its block label $t = b_j$, and we choose the new value $s$ with probability proportional to $e_{ts} + 1$ (this is not simply $e_{ts}$ to guarantee ergodicity, i.e. all values of $s$ can be chosen with nonzero probability). In other words, we inspect what is the typical block neighborhood of a neighbor $j$ to decide where the node $i$ is more likely to belong. This choice is particularly appealing since it is possible to sample the value of $s$ with a very simple and efficient algorithm. All we need is to write the move proposal probability as

$$p(r \to s | t) = \frac{e_{ts} + 1}{e_t + B} = (1 - R_t) \frac{e_{ts}}{e_t} + R_t \frac{1}{B}, \tag{8}$$

with $R_t = B/(e_t + B)$. Hence, in order to sample $s$ we proceed as follows: 1. A random neighbor $j$ of the node $i$ being moved is selected, and its block membership $t = b_j$ is obtained; 2. The value $s$ is randomly selected from all $B$ choices with equal probability; 3. With probability $R_t$ it is accepted; 4. If it is rejected, a randomly chosen edge adjacent to block $t$ is chosen, and the block label $s$ is taken from its opposite endpoint.

This algorithm is "rejection free" (despite step 3), since it always produces a value of $s$ with the desired probability after a single execution. It requires that a list of half-edges incident on each block is kept at all times. These lists can be updated efficiently in time $O(k_i)$ after the move of node $i$, since they do not need to be ordered, and incur a memory complexity of $O(E)$, so the whole approach is very easy to implement. The value of $s$ sampled this way still needs to be accepted or rejected depending on how it changes the entropy. In order for the steady state distribution of the Markov chain to be correct we still need to enforce detailed balance. Hence the final Metropolis-Hastings acceptance probability $a$ needs to be

$$a = \min \left\{ e^{-\beta \Delta \mathcal{S}_{t/c}} \frac{\sum_t p_t^i p(s \to r | t)}{\sum_t p_t^i p(r \to s | t)}, 1 \right\}, \tag{9}$$

where $p_t^i$ is the fraction of neighbours of node $i$ which belong to block $t$, and $p(s \to r | t)$ is computed after the proposed $r \to s$ move (i.e. with the new values of $e_{rt}$), whereas $p(r \to s | t)$ is computed before. As mentioned in the previous section, the computation of $\mathcal{S}_{t/c}$ can be done in $O(k_i)$ time, which is also the same complexity for the remaining terms of $a$. A full Monte Carlo sweep of the network can therefore be performed in $O(E)$ time.

If one chooses $\beta = 1$, the partitions are sampled with probability proportional to $e^{-\mathcal{S}_{t/c}}$, which correspond to the correct posterior probability that the fitted model matches the data. This can be useful in order to sample the marginal probability $\pi_r^i$ that node $i$ belongs to block $r$, which gives more detailed information on the network structure [5]. However, if one wants to minimize the description length, the ground state of $\mathcal{S}_{t/c}$ needs to be obtained via $\beta \to \infty$. This can be done by changing the value of $\beta$ either slowly (i.e. simulated annealing [6]) or abruptly (i.e. greedy minimization). The former approach avoids getting trapped local minima, but can be very slow and requires experimentation with the cooling schedule, while the latter is more efficient, but does not guarantee that the optimum is found. However, for sufficiently well-pronounced block structures, both approaches should produce comparable results. In the following examples we use a slightly less greedy version of the latter, simpler approach, which consists

in abruptly cooling the system but only once the Markov chain for $\beta = 1$ has been sufficiently equilibrated, but the overall method does not depend on how the limit $\beta \to \infty$ is eventually reached.

The equilibration criterion used was to keep track of the maximum and minimum values of $\Sigma_{t/c}$ and stop after $T$ successive sweeps occurred and both values did not change, where $T$ is made sufficiently high so that the results no longer depend on it. This criterion was applied twice in a row to get over "humps" in the value of $\Sigma_{t/c}$ when starting from a previously minimized state for a larger value of $B$ (see next section).

### A. Minimizing the description length $\Sigma_{t/c}$

Since the minimum value of $\mathcal{S}_{t/c}$ can be obtained independently for every value of $B$ with the above algorithm, the minimum value of $\Sigma_{t/c}$ (or equivalently $\Sigma_b$) can be obtained via a one-dimensional minimization on $B$. The most appropriate algorithm in this case is called golden search (a.k.a Fibonacci search) [7]. It consists in at first bracketing the minimum of $\Sigma_b$ by finding a triplet $(B_1, B_2, B_3)$, with $B_1 < B_2 < B_3$ such that $\Sigma_b|_{B=B_1} > \Sigma_b|_{B=B_2} < \Sigma_b|_{B=B_3}$. This is done easily by starting with $B_1 = 1$, $B_3 = B_{\max}$ and choosing $B_2 = B_3 - \lfloor B_3 - B_1 \rfloor_F$, where $\lfloor x \rfloor_F$ is the largest Fibonacci number smaller than $x$. This is repeated until the minimum is bracketed. After this, the intervals are progressively bisected with $B_2' = B_3' - \lfloor B_3' - B_1' \rfloor_F$, with $(B_1', B_3')$ being the largest of the intervals $(B_1, B_2)$ or $(B_2, B_3)$. Depending on the value of $\Sigma_b|_{B=B_2'}$, we choose the new interval as $(B_1|B_2, B_2', B_2|B_3)$ or $(B_1|B_2', B_2, B_2'|B_3)$, so that the new choice brackets the minimum. This algorithm contributes with a factor of $O(\ln B_{\max})$ to the overall complexity. The bisections chosen this way optimize the worse-case scenario, and guarantee that the global minimum is found as long as the function being minimized is convex. This needs not be the case for $\Sigma_b$ in general, so one can only guarantee that a local minimum is found. However, in most cases $\Sigma_b$ has an overall convex shape, even if not strictly so near the minimum, since it is exactly 0 for $B = 1$, less than zero for some $B > 1$ (if some block structure is detectable) and $\Sigma_b \to \infty$ for $B \to \infty$. If more precision is desired, one can perform the search for different initial ranges on $B$. Furthermore, since it is based on a minimization of $\mathcal{S}_{t/c}$ via Monte Carlo, it is often useful to perform multiple independent runs of the algorithm, and choose the best outcome.

A crucial part of the algorithm involves obtaining the minimum partitions for the different values of $B$ encountered during the bisections. A naïve application of the MC sweeps described above starting from a random partition would discard all the work done for the previous values of $B$. Instead, whenever minimizing $\Sigma_b$ for a given value of $B$, we start with the previously obtained solution for the smallest $B' > B$, and treat nodes belonging to the same blocks as single nodes, weighted according to $n_r$ and the edges between them weighted according to $e_{rs}$, such that the value of $\mathcal{S}_{t/c}$ is the same. The MCMC algorithm is performed for this graph until convergence, and afterwards the process is repeated with the original graph, so that nodes can be moved individually. This multilevel step can decrease significantly the mixing time of the Markov Chain at later steps.

The overall complexity of the entire algorithm is therefore $O(\tau E \ln B_{\max})$, where $\tau$ is the average mixing time of the Markov chain. If we have no prior information on $B_{\max}$, we need to assume $B_{\max} \sim \sqrt{E}$, in which case the complexity becomes $O(\tau E \ln E)$, or equivalently $O(\tau N \ln N)$ for sparse graphs with $N \sim E$.

An efficient C++ implementation of this algorithm is freely available as part of the graph-tool Python library at http://graph-tool.skewed.de.

## III. THE INTERNET MOVIE DATABASE (IMDB) NETWORK

The IMDB network was constructed by considering all available records in the IMDB database, available at http://www.imdb.com/interfaces. It contains comprehensive information of films, tv-shows and video games, and the cast of actors, as well as producers, directors, etc. Here we considered only the bipartite film-actor network, where a node represents either a film or an actor, and a given film is linked to its cast members. A 'film' designates any entry in the IMDB database, which can correspond to a theatrical release, as well as straight-to-video releases, tv-shows and even video games, and 'actor' designates any cast member. Extensive information is available for the films, including year of production, country of production, genres, and user supplied keywords. We have collected all available information ca. October 2012 into a network representing the full database. However, for many entries there is little to no additional information available, except the bare essentials such as title. Since we intend to interpret the overall large-scale structure, we pruned the database so that only entries with all of the mentioned metadata are included. Furthermore, we removed either actors which appear on only one film, and films with only one actor, since these entries only burden the analysis, without providing significant information on the overall network structure (this pruning was applied recursively, so what remained is the 2-core [8] of the network). The resulting network has $N = 372,787$ nodes ($275,805$ actors and $96,982$ films) and $E = 1,812,657$ edges (the average degree is hence $\langle k \rangle \approx 9.72$, and the actors appear on average on $\langle k \rangle_a \approx 6.57$ films, and the films have on average $\langle k \rangle_m \approx 18.7$ actors).

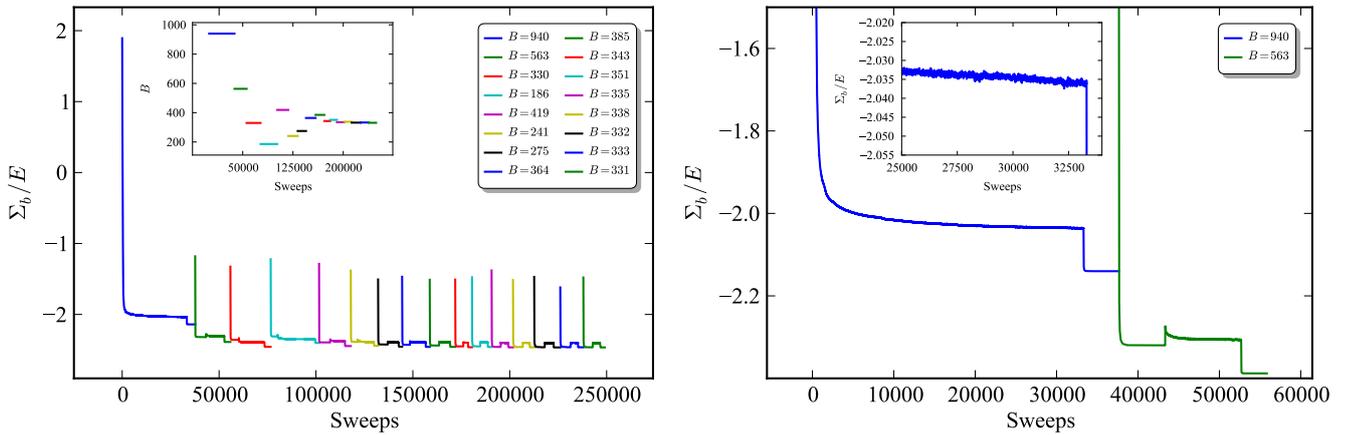

FIG. 1. *Left*: Typical run of the minimization algorithm for the IMDB network, for $T = 1000$. The inset shows the respective values of $B$. *Right*: Zoom in a specific region of the left figure. The first abrupt transition corresponds to the $\beta \to \infty$ cooling (the inset shows a further zoom in this region). The other two abrupt transitions for a smaller value of $B$ correspond to the switch from the multilevel step, and then finally the $\beta \to \infty$ cooling.

We applied multiple runs of the algorithm above on this network, and the partition with the minimum value of $\Sigma_b$ was obtained for $B = 332$. A typical run of the algorithm can be seen in Fig. 1. The obtained block structure can be seen in Fig. 2. The most obvious feature is that the block structure obtained is also bipartite, i.e. any given block is either only composed of actors or films. This feature arises out of the data itself, and is not *a priori* imposed. It does however often happen that the best partition contains a small minority of 2 or 3 blocks which contain both films and actors. This is due to films and actors which have a very small degree, and thus the bipartite nature cannot be easily detected. Nevertheless, the best partitions found (i.e. with the smallest value of $\Sigma_c$) corresponded to a fully bipartite structure.

Note that this type of fully dissortative block structure cannot be obtained with the more frequently used community detection methods, such as modularity optimization and many others, since they focus solely on the *opposite* case, where blocks form assortative connections.

More insight in the uncovered structure can be obtained by inspecting Fig. 3, which shows a graphical representation of the block structure, and Table I which shows the metadata which most often appears in each block. Without resorting to a more detailed correlation analysis, certain patterns are clearly recognizable. As mentioned in the main text, the film blocks are divided clearly according to the year and country of production, as well as genres. In Fig. 3 the layout is such that a rough time arrow pointing from bottom to top emerges. Films made in USA take a sizable portion of the graph, and are roughly divided in two main groups: Films made in the 20s-60s (a.k.a. the "Golden Age of Hollywood") and the more contemporary films from the 70s and onwards. In parallel one can distinguish films produced in European countries, such as UK, France, Italy and Germany, following a similar time line. Geographical and cultural similarity is also easily recognizable in Fig. 3, such as the proximity between British and American films, as well as Canadian and American ones, etc. In addition to this, films seem to be grouped further into genre classes. Although there is an abundance of seemingly nondescript "Drama" films, categories such as Animation, Western, Documentary, Action, Sport, Music and Adult films are clearly separated. In turn, the actors are grouped into blocks which seem strongly correlated with the film classes, but there are many actor blocks which are strongly connected to more than one film class. Looking more closely at specific well-known actors reveals intuitive patterns. As would be expected, actors which worked mostly together are confined to the same Block, such as the four "The Three Stooges" actors: Moe Howard, Larry Fine, Curly Howard, Shemp Howard, which are all confined to block 293, and most of their films to block 85. The same also holds for famous duos and groups such as Stan Laurel, Oliver Hardy (block 199), William Abbott and Lou Costello (block 231), Bud Spencer and Terrence Hill (block 253), and the Marx brothers, Chico Marx, Harpo Marx, Groucho Marx, Zeppo Marx (block 231), the Beatles, John Lennon, Paul McCartney, Ringo Starr, and George Harrison (block 274), and so on. Actors which have not worked together systematically, but made similar types of films or tv-shows in the same period of time also tend to be grouped together, such as contemporary comedians Mike Myers, Will Ferrel, Adam Sandler, Ben Stiller and Rob Schneider (block 243), and martial arts actors Bruce Lee and Jackie Chan (block 230). However, the strongest factor separating the actors seems to be geographical location and period of activity, rather than any other professional pattern. For instance, although all Monty Python members (John Cleese, Eric Idle, Terry Gilliam, Graham Chapman, Michael Palin, and Terry Jones) belong to the same block (314), they are accompanied by $\sim 1500$ other actors, including Sean Connery, Roger Moore, and Alec

Guinness, i.e. British actors active mostly during the 70s and 80s.

---

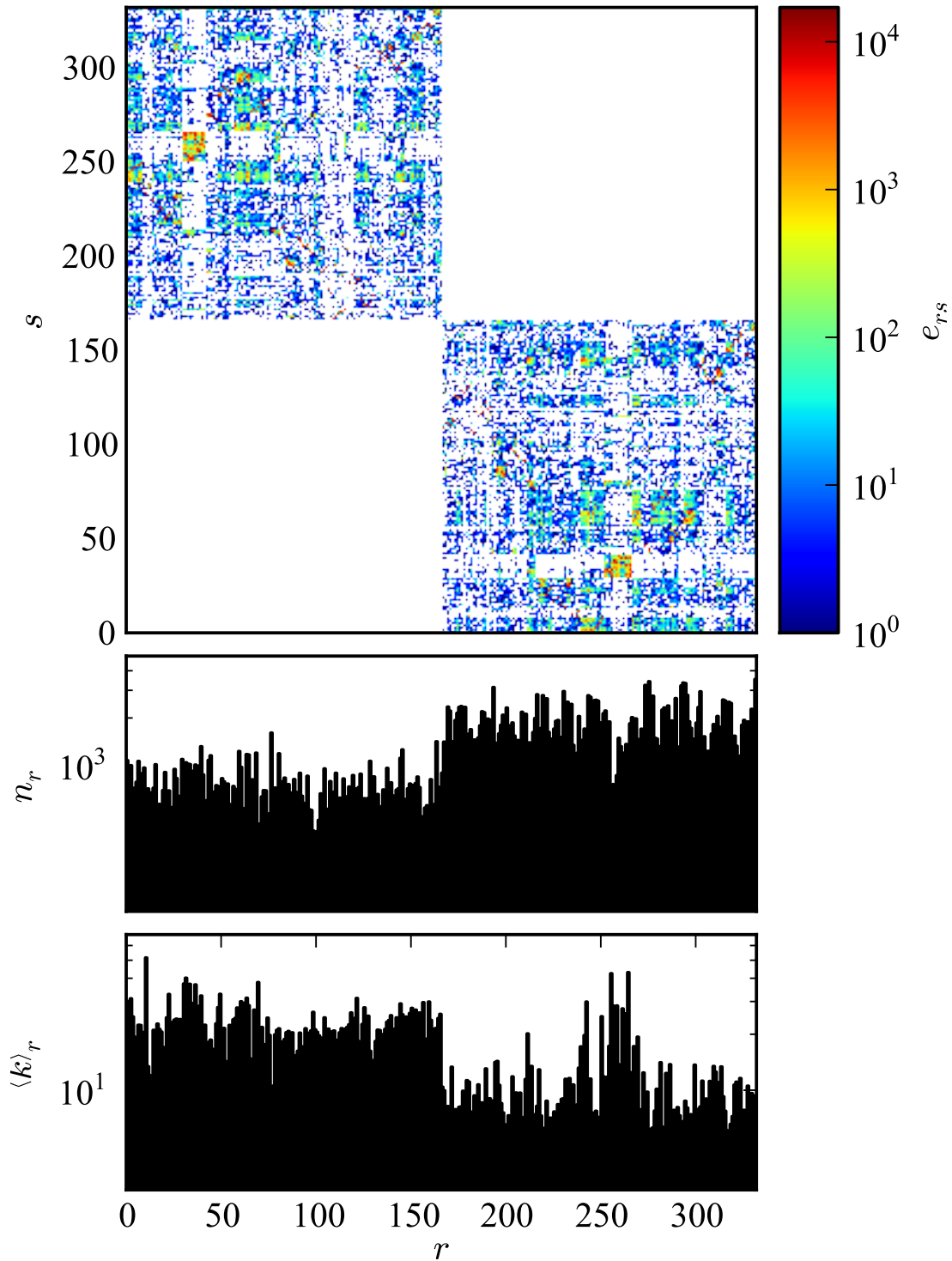

FIG. 2. Inferred block structure for the IMDB network for $B = 332$ which minimizes $\Sigma_c$, with a clear bipartite structure. The blocks in the range $[0, 164]$ are composed only of films, and the remaining 167 blocks only of actors. *Top:* The inferred $e_{rs}$ matrix, *Middle:* Block sizes $n_r$, *Bottom:* Average block degree $\langle k \rangle_r = e_r/n_r$.

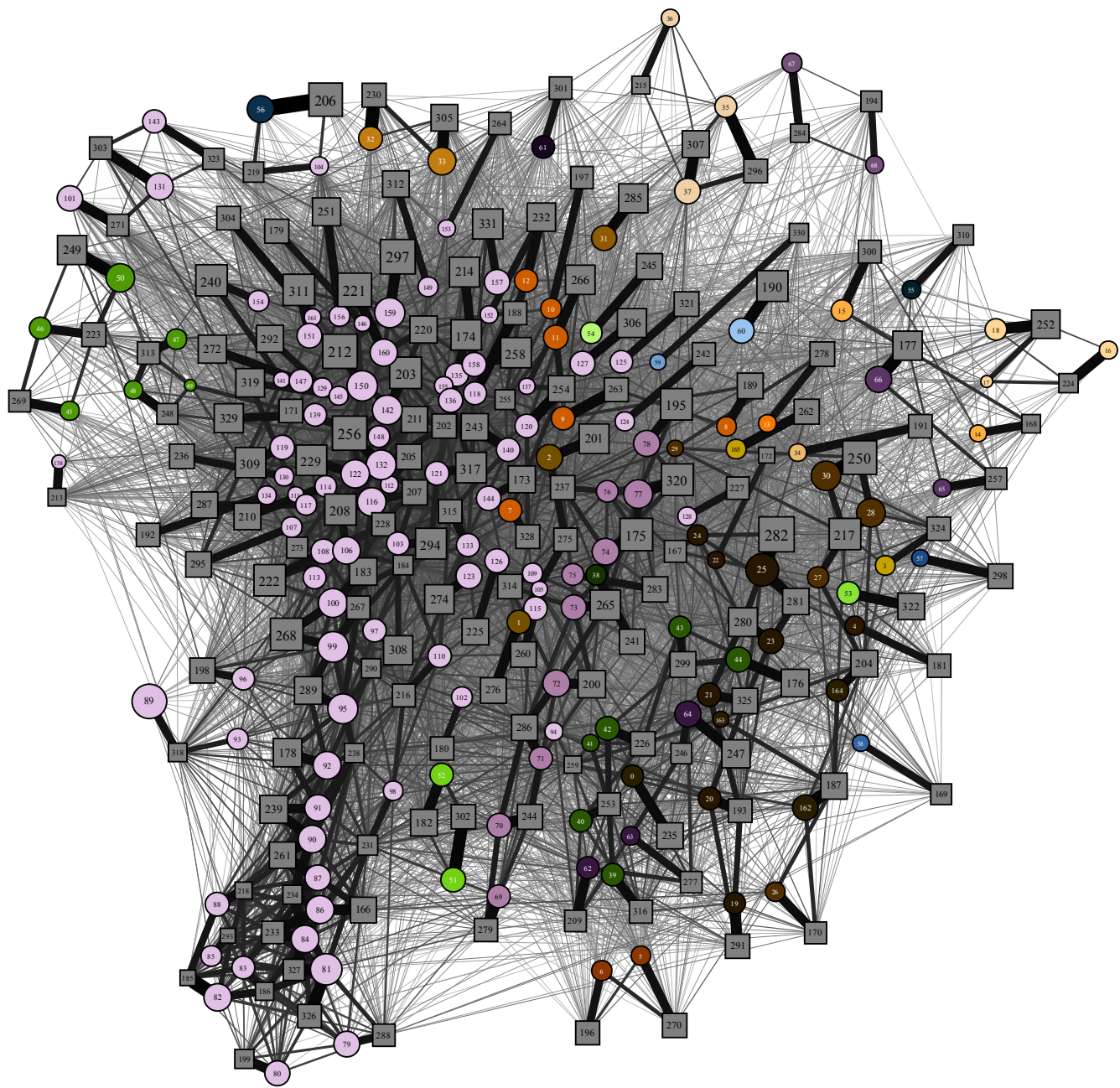
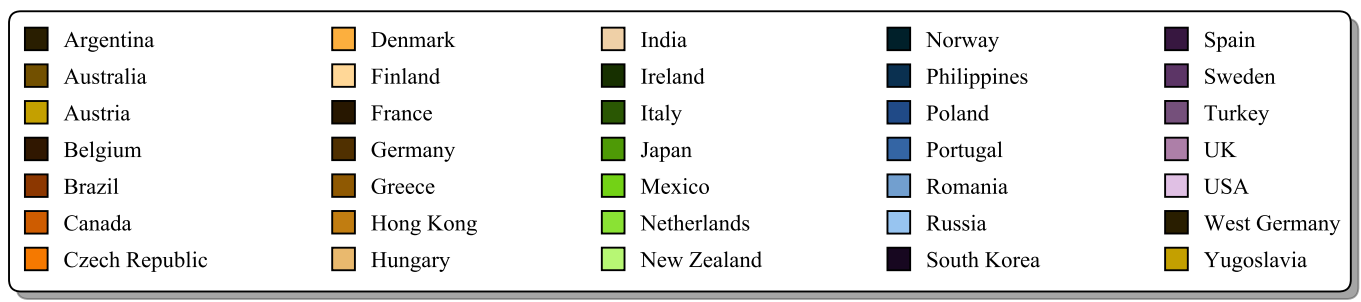

FIG. 3. Spring-block representation of the IMDB block structure of Fig. 2. Circles represent film blocks, and squares actor blocks. The node size and edge thickness correspond to the block sizes and number of edges between blocks. The color of the film blocks correspond to predominant country of production of the respective films, according to the legend. Further information on each block is given in Table I.

| Index | Year | Countries | Genres |
|---|---|---|---|
| 0 (140) | 1999 | **Argentina** | [Drama, Comedy] |
| 1 (117) | 1984 | **Australia** | [Drama, Comedy] |
| 2 (8) | 2003 | **Australia, USA** | [Drama, Comedy] |
| 3 (96) | 2001 | **Austria, Germany, Switzerland** | [Drama, Comedy] |
| 4 (161) | 1998 | **Belgium, France, Netherlands** | [Drama, Comedy] |
| 5 (131) | 1982 | **Brazil** | [Drama, Comedy] |
| 6 (134) | 1994 | **Brazil** | [Drama, Comedy] |
| 7 (14) | 1982 | **Canada, USA** | [Drama, Comedy] |
| 8 (97) | 1995 | **Canada** | [Drama, Comedy] |
| 9 (154) | 2001 | **Canada, USA** | [Drama, Thriller] |
| 10 (54) | 2002 | **Canada, USA** | [Drama, Comedy] |
| 11 (51) | 2005 | **Canada, USA** | [Drama, Comedy] |
| 12 (120) | 2005 | **Canada, USA** | [Drama, Thriller] |
| 13 (81) | 1994 | **Czech Republic, Czechoslovakia** | [Drama, Comedy] |
| 14 (108) | 1966 | **Denmark** | [Comedy, Family] |
| 15 (45) | 2002 | **Denmark** | [Drama, Comedy] |
| 16 (118) | 1949 | **Finland** | [Comedy, Drama] |
| 17 (113) | 1984 | **Finland** | [Drama, Comedy] |
| 18 (112) | 2000 | **Finland** | [Drama, Comedy] |
| 19 (127) | 1944 | **France** | [Drama, Comedy] |
| 20 (165) | 1963 | **France, Italy** | [Drama, Comedy] |
| 21 (164) | 1977 | **France, Italy** | [Drama, Comedy] |
| 22 (12) | 1988 | **France, UK** | [Adult, Horror] |
| 23 (160) | 1991 | **France** | [Drama, Comedy] |
| 24 (162) | 2000 | **France, USA, UK** | [Drama, Comedy] |
| 25 (163) | 2004 | **France** | [Drama, Comedy] |
| 26 (115) | 1935 | **Germany** | [Drama, Comedy] |
| 27 (42) | 1993 | **Germany, West Germany** | [Drama, Crime] |
| 28 (43) | 2000 | **Germany** | [Drama, Crime] |
| 29 (98) | 2001 | **Germany, USA, UK** | [Drama, Thriller] |
| 30 (44) | 2005 | **Germany** | [Drama, Comedy] |
| 31 (129) | 1994 | **Greece, USA** | [Drama, Comedy] |
| 32 (47) | 1977 | **Hong Kong, Taiwan** | [Action, Drama] |
| 33 (15) | 1999 | **Hong Kong, China** | [Drama, Action] |
| 34 (16) | 2000 | **Hungary, USA** | [Drama, Comedy] |
| 35 (116) | 1981 | **India** | [Drama, Action] |
| 36 (55) | 2002 | **India** | [Drama, Action] |
| 37 (119) | 2005 | **India** | [Drama, Romance] |
| 38 (17) | 2000 | **Ireland, UK, USA** | [Drama, Comedy] |
| 39 (109) | 1956 | **Italy, France** | [Drama, Comedy] |
| 40 (86) | 1969 | **Italy, France, Spain** | [Western, Comedy] |
| 41 (87) | 1972 | **Italy, France, West Germany, USA** | [Drama, Comedy] |
| 42 (85) | 1979 | **Italy, France** | [Comedy, Drama] |
| 43 (84) | 1993 | **Italy** | [Drama, Comedy] |
| 44 (83) | 2003 | **Italy** | [Drama, Comedy] |
| 45 (94) | 1961 | **Japan** | [Drama, Action] |
| 46 (92) | 1981 | **Japan** | [Drama, Action] |
| 47 (158) | 1994 | **Japan** | [Animation, Action] |
| 48 (159) | 2000 | **Japan** | [Animation, Action] |
| 49 (156) | 2003 | **Japan, USA** | [Action, Adventure] |
| 50 (128) | 2004 | **Japan** | [Drama, Comedy] |
| 51 (107) | 1961 | **Mexico** | [Drama, Comedy] |
| 52 (148) | 1996 | **Mexico, USA** | [Drama, Comedy] |
| 53 (19) | 1997 | **Netherlands** | [Drama, Comedy] |
| 54 (146) | 2000 | **New Zealand, USA, Canada** | [Drama, Comedy] |
| 55 (155) | 1997 | **Norway** | [Drama, Comedy] |
| 56 (104) | 1999 | **Philippines** | [Drama, Comedy] |
| 57 (147) | 2000 | **Poland** | [Drama, Comedy] |
| 58 (103) | 1994 | **Portugal, France** | [Drama, Comedy] |
| 59 (18) | 2001 | **Romania, USA** | [Drama, Horror] |
| 60 (50) | 1996 | **Russia, Soviet Union** | [Drama, Comedy] |
| 61 (13) | 1999 | **South Korea, Thailand, USA** | [Drama, Comedy] |
| 62 (132) | 1966 | **Spain, Italy** | [Drama, Comedy] |
| 63 (133) | 1987 | **Spain** | [Drama, Comedy] |
| 64 (91) | 2003 | **Spain** | [Drama, Comedy] |
| 65 (136) | 1962 | **Sweden** | [Drama, Comedy] |
| 66 (110) | 1995 | **Sweden** | [Drama, Comedy] |
| 67 (157) | 1967 | **Turkey** | [Drama, Romance] |
| 68 (105) | 2003 | **Turkey** | [Drama, Comedy] |
| 69 (135) | 1938 | **UK** | [Drama, Comedy] |
| 70 (139) | 1954 | **UK** | [Drama, Comedy] |
| 71 (138) | 1964 | **UK** | [Drama, Comedy] |
| 72 (137) | 1972 | **UK** | [Drama, Comedy] |
| 73 (26) | 1986 | **UK, USA** | [Drama, Comedy] |
| 74 (25) | 1997 | **UK, USA** | [Drama, Comedy] |
| 75 (27) | 2000 | **UK** | [Documentary, Comedy] |
| 76 (28) | 2002 | **UK, USA** | [Drama, Thriller] |
| 77 (29) | 2004 | **UK** | [Drama, Comedy] |
| 78 (20) | 2006 | **UK** | [Drama, Short] |
| 79 (40) | 1925 | **USA** | [Drama, Comedy] |
| 80 (114) | 1927 | **USA** | [Comedy, Short] |
| 81 (39) | 1934 | **USA** | [Drama, Comedy] |
| 82 (38) | 1938 | **USA** | [Western, Action] |
| 83 (35) | 1939 | **USA** | [Drama, Action] |
| 84 (36) | 1940 | **USA** | [Drama, Comedy] |
| 85 (41) | 1942 | **USA** | [Comedy, Short] |
| 86 (31) | 1944 | **USA** | [Drama, Comedy] |
| 87 (30) | 1948 | **USA** | [Drama, Comedy] |
| 88 (37) | 1949 | **USA** | [Western, Action] |
| 89 (76) | 1951 | **USA** | [Animation, Short] |
| 90 (33) | 1952 | **USA** | [Drama, Comedy] |
| 91 (34) | 1956 | **USA** | [Drama, Western] |
| 92 (32) | 1962 | **USA** | [Drama, Comedy] |
| 93 (77) | 1964 | **USA** | [Animation, Family] |
| 94 (101) | 1966 | **USA, UK** | [Drama, Comedy] |
| 95 (80) | 1970 | **USA** | [Drama, Comedy] |
| 96 (46) | 1971 | **USA** | [Drama, Comedy] |
| 97 (143) | 1973 | **USA** | [Drama, Comedy] |
| 98 (79) | 1975 | **USA** | [Documentary, Short] |
| 99 (145) | 1976 | **USA** | [Drama, Comedy] |
| 100 (144) | 1981 | **USA** | [Drama, Comedy] |
| 101 (130) | 1983 | **USA** | [Adult, Comedy] |
| 102 (93) | 1983 | **USA, Iran, Argentina** | [Drama, Comedy] |
| 103 (22) | 1985 | **USA** | [Drama, Comedy] |
| 104 (90) | 1988 | **USA, Philippines** | [Action, Drama] |
| 105 (99) | 1988 | **USA** | [Drama, Action] |
| 106 (2) | 1988 | **USA** | [Drama, Comedy] |
| 107 (3) | 1988 | **USA** | [Drama, Thriller] |
| 108 (1) | 1989 | **USA** | [Drama, Comedy] |
| 109 (24) | 1990 | **USA, UK** | [Documentary, Drama] |
| 110 (78) | 1991 | **USA** | [Documentary, Biography] |
| 111 (74) | 1992 | **USA** | [Drama, Comedy] |
| 112 (62) | 1993 | **USA** | [Drama, Comedy] |
| 113 (72) | 1993 | **USA** | [Animation, Family] |
| 114 (125) | 1994 | **USA** | [Drama, Comedy] |
| 115 (52) | 1994 | **USA, UK** | [Documentary, History] |
| 116 (0) | 1994 | **USA** | [Drama, Thriller] |
| 117 (7) | 1995 | **USA** | [Drama, Comedy] |
| 118 (142) | 1995 | **USA, Canada** | [Drama, Thriller] |
| 119 (5) | 1995 | **USA** | [Drama, Horror] |
| 120 (48) | 1996 | **USA, Canada** | [Drama, Thriller] |
| 121 (150) | 1996 | **USA** | [Drama, Comedy] |
| 122 (6) | 1996 | **USA** | [Action, Drama] |
| 123 (123) | 1996 | **USA, UK** | [Documentary, Music] |
| 124 (126) | 1997 | **USA, South Africa** | [Drama, Action] |
| 125 (89) | 1997 | **USA, Israel** | [Drama, Comedy] |
| 126 (53) | 1997 | **USA, UK** | [Documentary, Music] |
| 127 (88) | 1998 | **USA, Bulgaria, Indonesia** | [Drama, Comedy] |
| 128 (111) | 1998 | **USA, Chile, Soviet Union** | [Drama, Comedy] |
| 129 (151) | 1998 | **USA** | [Horror, Comedy] |
| 130 (23) | 1999 | **USA** | [Documentary, Drama] |
| 131 (9) | 1999 | **USA** | [Adult, Drama] |
| 132 (63) | 1999 | **USA** | [Drama, Comedy] |
| 133 (57) | 1999 | **USA** | [Documentary, Comedy] |
| 134 (141) | 2000 | **USA** | [Drama, Comedy] |
| 135 (4) | 2002 | **USA** | [Drama, Horror] |
| 136 (149) | 2002 | **USA** | [Drama, Comedy] |
| 137 (49) | 2003 | **USA, Canada** | [Drama, Comedy] |
| 138 (10) | 2003 | **USA** | [Sport, Action] |
| 139 (73) | 2004 | **USA** | [Drama, Comedy] |
| 140 (56) | 2004 | **USA** | [Music, Documentary] |
| 141 (67) | 2004 | **USA** | [Drama, Comedy] |
| 142 (64) | 2004 | **USA** | [Comedy, Drama] |
| 143 (11) | 2004 | **USA, France** | [Adult, Drama] |
| 144 (82) | 2004 | **USA** | [Documentary, Music] |
| 145 (75) | 2004 | **USA** | [Drama, Comedy] |
| 146 (21) | 2004 | **USA** | [Drama, Comedy] |
| 147 (122) | 2005 | **USA** | [Comedy, Drama] |
| 148 (71) | 2005 | **USA** | [Animation, Action] |
| 149 (61) | 2005 | **USA** | [Drama, Comedy] |
| 150 (59) | 2005 | **USA** | [Drama, Comedy] |
| 151 (66) | 2005 | **USA** | [Horror, Comedy] |
| 152 (121) | 2005 | **USA, Canada** | [Comedy, Drama] |
| 153 (70) | 2006 | **USA, Singapore** | [Drama, Comedy] |
| 154 (65) | 2006 | **USA** | [Comedy, Short] |
| 155 (58) | 2006 | **USA** | [Drama, Thriller] |
| 156 (124) | 2007 | **USA** | [Drama, Comedy] |
| 157 (153) | 2007 | **USA, Canada** | [Drama, Horror] |
| 158 (152) | 2007 | **USA** | [Drama, Comedy] |
| 159 (68) | 2008 | **USA** | [Short, Drama] |
| 160 (60) | 2008 | **USA** | [Comedy, Drama] |
| 161 (69) | 2008 | **USA** | [Drama, Comedy] |
| 162 (95) | 1958 | **West Germany, Austria** | [Comedy, Drama] |
| 163 (100) | 1969 | **West Germany, Italy, USA** | [Drama, Comedy] |
| 164 (102) | 1977 | **West Germany** | [Drama, Comedy] |

TABLE I. Aggregated metadata of the film blocks of Figs. 2 and 3. For each entry is given an unique index as labeled in Fig. 3; the ordering used in Fig. 2 in parentheses; the average year of production (rounded to the nearest integer); the countries of production, ordered according to frequency (countries which appear in $< 10\%$ of the films were omitted); and the two most frequent genres in the block.



\end{document}